\documentclass[12pt,a4paper]{article}
\usepackage{amssymb,amsfonts,epic,eepic,latexsym}
\makeatletter
\@addtoreset{equation}{section}
\makeatother
\def\rN{{\rm NS}}
\def\rR{{\rm R}}
\def\pu(#1){\Phi^{\sigma}(#1)}
\def\pd(#1){\Phi_{\sigma}(#1)}
\def\Hn{{\cal H}^{\rm NS}}
\def\Hr{{\cal H}^{\rm R}}
\def\H{\cal H}
\def\ze{\zeta}
\def\Z{{\bf Z}}
\def\Zh{\Z+{1\over 2}}
\def\tr{{\rm tr}}
\def\goto#1{\buildrel #1 \over \longrightarrow}
\def\n{\nonumber\\}
\def\a(#1){\alpha_{#1}}
\def\l{\sigma}
\def\Pu(#1,#2){\Phi^{#1}(#2)}
\def\Pd(#1,#2){\Phi_{#1}(#2)}
\def\z(#1){\zeta_{#1}}

\def\hh{{\rm H}}
\def\th{{\rm \Theta}}
\def\h1{{\rm H_1}}
\def\t1{{\rm \Theta_1}}

\begin{document}

\begin{titlepage}

\vspace*{\fill}

\begin{center}
{\large\bf Correlation functions for the Z-Invariant Ising model}
\vfill
{\sc J.R.Reyes Mart\'inez}\\[2em]

{\sl Institut f\"ur Theoretische Physik \\
     Freie Universit\"at Berlin \\
     Arnimallee 14, 14195 Berlin }

\vfill
{\bf Abstract}

\end{center}

\begin{quote}

The correlation functions of the Z-invariant Ising model are
calculated explicitely using the Vertex Operators
language developed by the Kyoto school.

\end{quote}
\noindent{Sept 1996}
\end{titlepage}

\paragraph{Introduction}

In the last few years the Kyoto school has made a breakthrough in
the calculation of correlation functions and form factors
of Integrable Models
on the lattice by treating them directly in the thermodynamic
limit. This was possible because 
the physical quantities (observables and space of states) could
be translated into the powerful language of representation theory
of affine quantum algebras. Fundamental for this progress was
an understanding of the combinatorial aspects of the calculation
of order parameters  through Baxter's
corner transfer matrix formalism. The next step was the interpretation
of the CTM Hamiltonian as the derivation generator in the quantum affine
algebra. Finally the complete apparatus of representation theory
developed in \cite{FR} was employed with the introduction of
intertwining operators for different representations - the so-called
Vertex Operators.

The success of this method has stimulated the
search for new deformed affine-algebras  and their representation
theory. It has also been possible to calculate correlation
functions without the knowledge of the dynamical symmetry \cite{JM}
in a way similar to the bootstrap program for form-factors 
in integrable field theory. Interesting also is the development
in \cite{PL} where techniques of Conformal Field Theory were
directly 'deformed' to the setting of RSOS models. This progress
proves interesting not only in understanding the integrability
of lattice models but also around in understanding
the issue of 'quantum symmetry' in CFT where up to know the
concept of quantum group has not been of much use and where
the more abstract methods of algebraic quantum field theory have
yet to  ripen. For this reason the study of the simplest integrable
model, the Ising model, with this new method proves to be 
a good laboratory for understanding rigorously the new methods
independent of the setting of quantum algebras.

In this letter we restrict to the computation of physical quantities
and show how the methods 
developed in \cite{FJMMN} are surprisingly simple in 
obtaining correlation functions that by the old transfer matrix methods
are technically intricate. In the above paper it was shown how to
calculate the nearest diagonal order-order and disorder-disorder
correlations functions; we extend this without further pain
to a general distance and show by using a theta functions identity
the validity of an old conjecture of Baxter \cite{spin}, but
we can go further  and integrate this equation obtaining
an explicit expression for the two-point correlation functions.
In the last paragraph we make some comments on futher work possible
using the methods of this letter.

\paragraph{Vertex Operators}

We restrict to the ferromagnetic Ising model 
in the edge description,
below the critical temperature i.e. with 
$0< k < 1$,  $k = (\sinh2K\sinh2L)^{-1}$. Here $K>0$ and
$L>0$ are the coupling 
constants in the two orthogonal directions and
we use Onsanger's elliptic parameterization:
$\sinh(2K)=-i{\rm sn}(iu)$, $\sinh(2L)=i/k{\rm sn}(iu)$ where
${\rm sn}(u)$ is a Jacobian  elliptic
function with half-periods  $I$, $iI'$, $k$ is the
modulus, and $0< u< I'$ (see e.g. \cite{BaxBk}). We will also
employ the multiplicative 
parameters $\ze := \exp(-\pi u/2 I)$ and
$x :=\exp(-\pi I'/2 I)$).

Viewing now the two-dimensional lattice in the diagonal
and dividing
the lattice in four equal quadrants to apply
Baxter's CTM method we have two cases according
to if the central point belongs or not to the lattice;
this defines  {\em two} CTM's that act on the
two sectors of the model, $\rN$ and $\rR$,
respectively (in analogy to CFT).
Diagonalising the adjoint action of the CTM Hamiltonians 
we can describe the two sectors as Fock spaces (called
$\Hn$ and $\Hr$)  of
two irreducible representations of free fermions.
We have:
\begin{eqnarray*}
&&[D^\rN,\phi^\rN_r]=-2r\phi^\rN_r,\quad
[D^\rR,\phi^\rR_r]=-2r\phi^\rR_r ,
\end{eqnarray*}
where 
$r\in\Zh$ for the $NS$ sector,
$r\in\Z$ for the $R$ sector and
$D^\rN$, $D^\rR$ are the respective CTM hamiltonians.

There are two vertex operators (VOs in the following)
depending on which sector they intertwine:
\begin{eqnarray*}
&&\Phi_{\rN}^{\rR}(\zeta):\Hn\goto{} \Hr, \\
&&\Phi_{\rR}^{\rN}(\zeta):\Hr\goto{} \Hn,
\end{eqnarray*}
accordingly they intertwine the above fermions in a manner which
can be written in the surprising elegant way as follows:
\begin{eqnarray}
\label{eqn:inter1}
\phi^\rN(z)\Phi^\sigma(\zeta)&=&
f(z\zeta^2)\Phi^{-\sigma}(\zeta)\phi^\rR(z), \n
\phi^\rR(z)\Phi_\sigma(\zeta)&=&
f(z\zeta^2)\Phi_{-\sigma}(\zeta)\phi^\rN(z), 
\end{eqnarray}
\begin{eqnarray}
\label{eqn:inter1.2}
\sigma \Phi^\sigma(\zeta)&=&\Phi^\sigma(\zeta)\psi_1^\rR(\zeta), \n
\sigma \Phi_\sigma(\zeta)&=&-i\Phi_{-\sigma}(\zeta)\psi_1^\rN(\zeta).
\end{eqnarray}
Here we use
the simplified notation of \cite{FJMMN} 
$\Phi^{\rN,\sigma}_\rR(\zeta)=\pu(\ze)$,
$\sigma$ being equal to $+1$ or $-1$ distinguishes the
two sub-sectors (with even or odd parity)
of the $\rN$ sector and
$\Phi_{\rN,\sigma}^{\rR}(\zeta)=\pd(\ze)$, also we define
$[z] := f(z)= -\sqrt{k}{\rm sn}(v)$ with $z=\exp(i\pi v/I)$
and we have introduced the generating functions
\begin{eqnarray}
\label{eqn:phi}
&&\phi^\rN(z)=\sum_{r\in \Zh}\phi^\rN_r z^{-r},\quad 
\phi^\rR(z)=\sum_{s\in \Z}\phi^\rR_s z^{-s},\n 
&&[\phi^\rN(z),\phi^\rN(w)]_+ =
\delta^{\rN}(x^2z/w)+ \delta^{\rN}(x^{-2}z/w),\n
&&[\phi^\rR(z),\phi^\rR(w)]_+ =
\delta^{\rR}(x^2z/w)+ \delta^{\rR}(x^{-2}z/w),\n
&&\delta^{\rN}(z):=\sum_{r\in\Zh}\left(\frac{z}{w}\right)^r,\quad
\delta^{\rR}(z):=\sum_{r\in\Z}\left(\frac{z}{w}\right)^r,
\end{eqnarray}
and
\begin{eqnarray*}
\psi_1^\rN(\zeta)&=&
\oint{dz\over 2\pi i z} [z]_{\rN}\phi^\rN(z/\zeta^2),\\
\psi_1^\rR(\zeta)&=&
\oint{dz\over 2\pi i z} [z]_{\rR}\phi^\rR(z/\zeta^2).
\end{eqnarray*}
where
\begin{eqnarray*}
&&[z]_{\rN}=\sqrt{2I k\over \pi}{\rm cn}(v)
=\sqrt{2\pi \over kI}\sum_{r\in \Zh}\eta^{-1}_r z^r,\n
&&[z]_{\rR}=\sqrt{ 2I\over \pi}{\rm dn}(v)
=\sqrt{2\pi \over I}\sum_{s\in \Z}\eta^{-1}_s z^s, \n
&&\eta_r =x^{2r} + x^{-2r}.
\end{eqnarray*}
By crossing symmetry i.e. $u\rightarrow I'-u$ 
or in the multiplicative language $\ze\rightarrow x\ze$
we can obtain  (using also the transformation
$f(x^2z)=1/f(z)$) similar relations with the parameter
of the VO replaced with $x\ze$.
Finally the VOs satisfy the following unitary relations:
\begin{eqnarray}
&&\sum_\sigma
\Phi_{\sigma}(x\zeta)
\Phi^\sigma(\zeta)=g^{\rm R}\times {\rm id}_{\H^\rR},
\label{eqn:inverse1}\\
&&
\Phi^{\sigma}(x\zeta)\Phi_{\sigma}(\zeta)
=g^{\rm NS}\times {\rm id}_{\H^{\rN,\sigma}},
\label{eqn:inverse2}
\end{eqnarray}
where the scalars $g^{\rm R}$ and $g^{\rm NS}$ are given by
\begin{eqnarray*}
&&g^{\rm R}=
{(x^4;x^4,x^8)_\infty^2 \over (x^2;x^4,x^4)_\infty},\qquad
g^{\rm NS}=
{(x^8;x^4,x^8)_\infty^2 \over (x^6;x^4,x^4)_\infty}.
\end{eqnarray*}
For the infinite product notations see e.g. \cite{FJMMN}.
\paragraph{Two-point correlation functions}
We can now calculate the correlation functions (CF) for
the model with an arbitrary number of inhomogeneities.
As explained in \cite{FJMMN} the CF
are giving as traces of the VOs, in an Appendix of the same
paper the nearest order-order and disorder-disorder
CF were obtained. The generalization to arbitrary
separation is immediate. Let us call them
$g^{\sigma}_{2n}(\a(1),\a(2),\ldots,\a(2n))$ and
$g^{\mu}_{2n}(\a(1),\a(2),\ldots,\a(2n))$ respectively (when we
don't want to emphasize the spectral parameters $g^\l_{2n}$ and
$g^\mu_{2n}$ simply),
where $\ze_j=\exp(i\pi\a(j)/2I)$,  then we have:
\begin{eqnarray}
  \label{eqn:order}
g^{\sigma}_{2n}(\a(1),\a(2),\ldots,\a(2n))=
\frac{\sum_{\l;\mu_i;\l'}\l\l'\tr_{\rN}\left(x^{2D^{\rN}}
F(\l,\l';\mu_i;\z(1),\ldots,\z(2n))\right)}
{(g^{\rN}g^{\rR})^n\tr_{\rN}\left( x^{2D^{\rN}}\right)},
\end{eqnarray}
\begin{eqnarray*}
&&F(\l,\l';\mu_i;\z(1),\ldots,\z(2n))= \n
&&\Pu(\l',x\z(2n))\Pd(\mu_{n-1},x\z(2n-1))
\cdots\Pu(\mu_{1},x\z(2))\Pd(\l,x\z(1))\Pu(\l,\z(1))\Pd(\mu_{1},\z(2))
\cdots \Pu(\mu_{n-1},\z(2n-1))\Pd(\l',\z(2n)),
\end{eqnarray*}
where the subscript $\mu_i$ in the sum above means the sum for all
$1\leq i\leq n-1$. Similarly for the disorder-disorder case:
\begin{eqnarray}
  \label{eqn:disorder}
g^{\mu}_{2n}(\a(1),\a(2),\ldots,\a(2n))=
\frac{\sum_{\mu_i;}\tr_{\rR}\left(x^{2D^{\rR}}
G(\mu_i;\z(1),\ldots,\z(2n))\right)}
{(g^{\rN}g^{\rR})^n\tr_{\rR}\left( x^{2D^{\rR}}\right)},
\end{eqnarray}
\begin{eqnarray*}
&&G(\mu_i;\z(1),\ldots,\z(2n))= \n
&&\Pd(\mu_{n},x\z(2n))\Pu(\mu_{n},x\z(2n-1))
\cdots\Pd(\mu_1,x\z(2))\Pu(\mu_1,x\z(1))
\Pd(-\mu_1,\z(1))\Pu(-\mu_{1},\z(2))
\cdots \Pd(-\mu_{n},\z(2n-1))\Pu(-\mu_n,\z(2n)),
\end{eqnarray*}
with $1\leq i\leq n$. To obtain explicit expressions for the 
above we need the following equalities that can be proved
using eqs. (\ref{eqn:inter1}),(\ref{eqn:inter1.2}), 
(\ref{eqn:inverse1}) and (\ref{eqn:inverse2}) 
(see also \cite{FJMMN}):
\begin{eqnarray}
\label{eqn:4d}
\sum_{\l_j}\Phi_{\l_j}(x\ze_{l})\Phi^{\l_j}(x\ze_{i})
\Phi_{-\l_j}(\ze_i)\Phi^{-\l_j}(\ze_{l})= 
ig^{\rm NS}g^{\rm R}\oint_{w_{l}}\oint_{w_i}
[w_{l} l]_{\rm R}[w_i i]_{\rm NS}[w_i l]
\phi^\rR(w_{l})\phi^\rR(w_i),
\end{eqnarray}
\begin{eqnarray}
\label{eqn:2d}
&&\sum_{\l_1}\l_1\Phi_{\l_1}(x\ze_1)\Phi^{\l_1}(\ze_1)=
g^\rR\oint_{w_1}[w_11]_{\rR}\phi^\rR(w_1),\n
&&\sum_{\l_{2n}}\l_{2n}
\Phi^{-\l_{2n}}(x\ze_{2n})\Phi_{\l_{2n}}(\ze_{2n})=
-i g^\rN\oint_{w_{2n}}[w_{2n}2n]_{\rN}\phi^\rN(w_{2n}),
\end{eqnarray}
where we changed the variables to $w_i = z_i/\ze^2_i$ and used
the notation $\oint_i =\oint\frac{dw_i}{2\pi i w_i}$,
$[w_ij]=[w_i\ze^2_j]$. Because of eq. (\ref{eqn:phi}) the
fermions in expression (\ref{eqn:4d}) can be anticommutated
with only a sign change. In the case of $g^\mu_{2n}$ we need only
relation (\ref{eqn:4d}); starting at the center in the numerator
of expression (\ref{eqn:disorder}) we interchange each time
four VOs with to fermions, these we intertwine to the extreme
left or right using (\ref{eqn:inter1}) paying attention          
to the change of representations we do this successively n times
to arrive at:
\begin{eqnarray*}
 (-ig^{\rm NS}g^{\rm R})^{n}\prod_{1\leq i \leq 2n}\oint_{w_i}
[w_1]_{\rN}[w_2]_{\rR}\cdots
[w_{2n-1}]_{\rN}[w_{2n}]_{\rR}\times\!\! 
\prod_{1\leq i < j \leq 2n}[w_ij]\prod_{i=1}^{2n}\phi^\rR(w_i),
\end{eqnarray*}
where we have $2n$ integrations and the product of fermions
is ordered from left to right with increasing indices.
Similarly for the $g^\l_{2n}$ but starting with the first eq.
in (\ref{eqn:2d}) and ending with the second one,
the final expression is obtained interchanging the $\rR$ with
$\rN$ in the expression above. At the end we are left with,
using the notation $\delta^\rR(ij)=\delta^\rR(x^2w_i/w_j)$, 
\begin{eqnarray*}
&&{\tr_{\Hr}\left(x^{2D^\rR}\phi^\rR(z_1)\phi^\rR(z_2)\cdots
\phi^\rR(z_{2n-1})\phi^\rR(z_{2n})\right)
\over \tr_{\Hr}\bigl(x^{2D^\rR}\bigr)}=\! 
\sum\limits_{p}{}^{'}
\!\varepsilon_p\delta^\rR(p_1p_2)\delta^\rR(p_3p_4)\cdots
\delta^\rR(p_{_{2n-1}}p_{_{2n}})\!.
\end{eqnarray*}
Here $p_1,\ldots, p_{2n}$ is some permutation of the numbers
$1,2,\ldots ,2n$, $\sum_p'$ 
is a summation over all $(2n-1)!!$ permutations which
satisfy the restrictions 
$p_{2m-1} < p_{2m}$ for $1<m<n$ and $p_{2m-1}
< p_{2m+1}$ for $1<m<n-1$ and 
$\varepsilon_p$ is  the parity of the permutation.
There is a similar expression 
for the $\rN$ case (note that the integrations 
in this case are well defined although the function $\delta^\rN(ij)$
has half-integral powers in the integration variable $z$) .

Before inserting this formula under the integral
expression it is more convenient to antisymmetrize the factor
multiplying the fermions in all the $w_i$ variables (under the
integrals the fermions can be anticommutated as already observed),
for this we change the spectral 
parameters and integration variables
to their additive version,
\begin{eqnarray*}
&&\oint_{w_i}\rightarrow \int_{-I}^{I}{dv_i\over 2I},\hspace{3cm}
\delta^{\rN,\rR}(ij)\rightarrow 2I
\delta^{\rN,\rR}(v_i-v_j+iI'), \\
&&[w_ii]_{\rN}=\sqrt{2Ik'\over\pi}{\h1(v_i-i\a(i))
\over\th(v_i-i\a(i))},
\quad[w_ii]_{\rR}=\sqrt{2Ik'\over\pi}{\t1(v_i-i\a(i))
\over\th(v_i-i\a(i))},\\
&&[w_ij]     = -{\hh(v_i-i\a(j))\over\th(v_i-i\a(j))},
\end{eqnarray*}
where we introduced  the Jacobi theta
functions $\hh$,$\th$,$\h1$ and $\t1$ (see e.g.\cite{grad}).
The result of the antisymmetrization is:
\begin{eqnarray}
\label{eqn:theta}
&&{\prod_{i=1}^n\h1(v_{2i-1}-i\a(2i-1))\t1(v_{2i}-i\a(2i))
\prod_{1\leq i < j \leq 2n}\hh(v_i-i\a(j))\over 
\prod_{1\leq i \leq j \leq 2n}\th(v_i-i\a(j))}+ \mbox{perm} = \n
&&{\prod_{1\leq i < j \leq 2n}\th(i\a(i) - i\a(j))\hh(v_i-v_j)
\left(\h1(0)\t1(0)\right)^n \over \h1(0)\prod_{i=1}^{2n}
\prod_{j=1}^{2n}
\th(v_i-i\a(j))}\!\times\!
\h1\!\!\left(\sum_{i=1}^{2n}v_i-i\sum_{r=1}^{2n}
\a(r)\!\right),
\end{eqnarray}
where by perm we mean the antisymmetrization in all $v_i$. The same
formula is valid under the interchange of $\t1$ with $\h1$. This
identity is proved showing first that both sides transform
identically under the periods $2I$ and $2iI'$ in all $v_i$, and then
through an inductive process it is possible show that  
the residues of the simple poles
at e.g. $v_1=i\a(1)+iI'$ are identical for the two sides.
As a simple consequence of Liouville's theorem the rhs can only
differ by an additive constant, which is proved to be zero
by comparing the two sides at e.g. $v_{2n}=i\a(2n)$.
Substituting all these expressions and integrating out the 
delta functions, we obtain:
\begin{eqnarray}
  \label{eqn:baxt}
&&g_{2n}^{\mu}(\a(1),\cdots ,\a(2n))= \n
&&{1\over n!\h1(0)}{\Big({-k^{\frac{1}{2}}\th^3(0)\over
2\pi}\Big)^n}
\prod_{i=1}^{n}\int_{-I}^{I}\!\!\! dv_i 
{\prod\limits_{1\leq r < s\leq 2n}\th(i\a(r)-i\a(s))
\prod\limits_{1\leq j< k\leq n}
h^2(v_j-v_k)
\over \prod_{r=1}^{2n}\prod_{j=1}^{n}h(v_j-i\a(r))}\times  \n
&&\hspace{8cm}\times\Psi_n(2\sum_{j=1}^nv_j - i\sum_{r=1}^{2n}\a(r)),
\end{eqnarray}
\begin{eqnarray*}
\Psi_n(u) = \left\{ \begin{array}{cc} 
\h1(u) \qquad \mbox{if}\qquad n\qquad \mbox{ even} \\
\t1(u) \qquad \mbox{if}\qquad n\qquad \mbox{ odd}
\end{array} \right.
\end{eqnarray*}
with $h(u) = \th(u)\hh(u)$. This formula proves
Baxter's conjecture in \cite{spin} (the formula was
proved in the critical case in \cite{AP}).
We obtain $g_{2n}^\l$ 
interchanging $\t1$ with $\h1$ in (\ref{eqn:baxt}).

We can go further and combine the above theta function identity with 
the following one (this is a shifted version of Fay's identity,
see e.g. \cite{Fay}):
\begin{eqnarray*}
\det\limits_{_{1\leq i< j\leq 2n}}
\left[\frac{\h1(v_i-i\a(j))\hh(0)'}{\th(v_i-i\a(j))\t1(0)}\right]=
\frac{\left(\hh'(0)\right)^{2n}}{\h1(0)}
\frac{\prod\limits_{i<j}\hh(v_i-v_j)
\hh(i\a(i)-i\a(j))}{\prod\limits_{i,j}\th(v_i-i\a(j))}
\h1\!\left(\!\sum_i(v_i-i\a(i))\!\right)
\end{eqnarray*}
and another identity interchanging $\h1$ with $\t1$.
We see that multiplying the rhs of the above eq. by appropriate
constants terms we obtain the rhs of (\ref{eqn:theta}), expanding
the determinant we have reduced the problem to independent
known quadratures,
collecting all constants we arrive at the final result:
\begin{eqnarray}
  \label{eqn:two}
&&g^{\mu,\l}_{2n}=\frac{{\rm Pfaffian}
\left(h^{\mu,\l}(\a(i)-\a(j))\right)}
  {{\rm Pfaffian}\left(\sqrt{k}{\rm sn}(i\a(i)-i\a(j))\right)},\\
&&h^\mu(\alpha):= 
-\frac{2I}{\pi}\left(\frac{k'}{k}\right)^{\frac{1}{2}}
\frac{\t1'(i\alpha)}{\th(i\alpha)},\quad
h^\l(\alpha):=
-\frac{2I}{\pi}(k')^{\frac{1}{2}}
\frac{\h1'(i\alpha)}{\th(i\alpha)}.
\end{eqnarray}
Here we also employed a lemma proved in \cite{PT} (eq. (5.8)):
\begin{eqnarray*}
{\rm Pfaffian}\left(\sqrt{k}{\rm sn}(i\a(i)-i\a(j))\right)=
\prod\limits_{1\leq i< j\leq 2n}
\left(\sqrt{k}{\rm sn}(i\a(i)-i\a(j))\right).
\end{eqnarray*}

Expression (\ref{eqn:two}) is symmetric in all spectral parameters
and satisfies the recurrence relation of Baxter
\begin{eqnarray*}
&&g^{\mu,\l}_{2n}(\a(1),\a(2),\cdots,\a(2n-1),\a(2n-1)+ I')= 
g^{\mu,\l}_{2n-2}(\a(1),\a(2),\cdots,\a(2n-2)),\n
&&g^{\mu,\l}_2(\a(1),\a(1)+I')=1.
\end{eqnarray*}
As a final check of the expression for the correlation functions
we calculated the homogeneous limit, when all spectral parameters
are equal of the $g^{\mu}_4$ correlation function, taking the 
necessary limits  we obtain after a tedious calculation:
\begin{eqnarray*}
g^{\mu}_{4,homo.}= \frac{4}{3\pi^2 k^2}\big[
(5-k^2)E(k)^2 + 8(k^2-1)E(k)I(k)+
3(k^2-1)^2I(k)^2\big]
\end{eqnarray*}
where $I(k)$, $E(k)$ are the complete elliptic integrals of the
first and second kinds. This result coincides with the value
obtained by Toeplitz determinant method (see \cite{ghosh}).

Employing similar methods as for the two-point functions
we can write immediately
the higher CF as a multiple integral expression (see \cite{me}).

\paragraph{Conclusions}

We  have shown that treating the Ising model
at the thermodynamic limit simplifies considerably the
calculations of physical quantities (we remark that 
using the full Z-invariance of the model gives us simpler
results than the homogeneous case).
It might also be
possible to use the theta function identities 
to explicitly integrate the correlation functions
of the six-vertex model obtained in \cite{six-vertex}
(see also \cite{six-vertex-book}) through bosonization. 
Using the
commutation relations of the VO and its unitary properties
we can write difference equations, it will be interesting
to understand in this way the CF ,
this is important to understand the eight-vertex model
CF out of its decoupling point (where it decomposes
in two Ising models) where we cannot use the translation
to fermions but the problem can also be formulated as the solutions
of some difference equations as in the form-factor program
of integrable field theory. The simplicity of the
result (\ref{eqn:two}) gives some hope in the possibility of 
generalizing this result for the eight-vertex model
(by some deformation with an extra parameter).
Further work is also suggested by
\cite{PL} where a deformed energy tensor field plays
a r\^ole of dynamical symmetry for the RSOS models, the Ising
being the simplest one in this category it will be interesting
to understand in this context this symmetry and its connection
or not with the temperature-like states that happen to describe 
the CF of the theory and the toroidal parameter space of the
elliptic functions used (see \cite{IT}). 

\paragraph{Acknowledgments}

I thank Profs. B.Davies and I.Peschel for conversations.
I want also to thank Prof. J. Perk for pointing out
that he and H. Au-Yang obtained similar results using
another method, but due to a computer accident they lost their
calculations. Prof. J. Perk has also pointed to me reference 
\cite{Yamada} (and references therein to earlier papers 
by the same author) where similar results are given.


\end{document}